\documentclass[]{interact}

\usepackage{xspace}
\usepackage{epstopdf}
\usepackage[caption=false]{subfig}
\usepackage[dvipsnames]{xcolor}
\usepackage{color,soul}
\usepackage{textcomp}
\usepackage[export]{adjustbox}
\usepackage{setspace}
\usepackage[none]{hyphenat}

\newcommand{\wn}{cm$^{-1}$\xspace}
\newcommand{\PI}{H$_2$CCCH$^+$\xspace}
\newcommand{\propenylcat}{C$_3$H$_3^+$\xspace} 
\newcommand{\propenylidene}{C$_3$H$_2$\xspace} 
\newcommand{\prop}{C$_3$H\xspace} 

\begin{document}

\articletype{Special Issue}

\title{High Resolution Rovibrational and Rotational  Spectroscopy of  \PI}

\author{
\name{Weslley Guilherme Dias de Paiva Silva\textsuperscript{a}, Divita Gupta\textsuperscript{a}, Eline Plaar\textsuperscript{a}, Jos\'e Luis Dom\'enech\textsuperscript{b}, Stephan Schlemmer\textsuperscript{a}, Oskar Asvany\textsuperscript{a}\thanks{CONTACT:   Oskar Asvany (asvany@ph1.uni-koeln.de)}}
\affil{\textsuperscript{a}
I.  Physikalisches Institut, Universit\"at zu K\"oln, Z\"ulpicher Str.~77, 50937 K\"oln, Germany;\\
\textsuperscript{b}
Instituto de Estructura de la Materia, IEM-CSIC, Serrano 123, 28006 Madrid, Spain}
}

\maketitle

\begin{abstract}
The rovibrational spectrum of the molecular ion \PI was investigated in a 
4~K cryogenic ion trap instrument employing the  leak-out spectroscopy method. 
Transitions within the fundamental $\nu_1$ (C-H stretch) and the 
combination band $\nu_3$+$\nu_5$ (C-C stretches) were  
detected, the search aided by high level quantum chemical calculations. 
The analysis of the rovibrational measurements enabled us to predict the rotational structure of the ground state. 
Using a rotational-vibrational double-resonance scheme, 
14 pure rotational transitions were measured.
This, in turn, 
led to the radio astronomical detection of \PI in the interstellar medium, 
as recently reported (Silva et al., Astron. Astrophys. 676, L1, 2023). 
\end{abstract}



\section{Introduction}

Astrochemical models (see, e.g., \cite{sip16}) show that \propenylcat cations play an important role in the chemical evolution of the interstellar medium (ISM). They
are considered to undergo  dissociative recombination  with electrons,
leading to the formation of neutral \propenylidene and \prop, 
which are well known molecules in space \cite{Thaddeus1985_c3h,Thaddeus1985_c3h2,Yamamoto1987,Cernicharo1991}.
The two most stable isomers of \propenylcat are the cyclopropenyl (\emph{c}-\propenylcat, \emph{D}$_3$$_h$) 
and the open chain propargyl (\PI, \emph{C}$_2$$_v$) cation. 
Although \PI is 27.9~kcal/mol less stable than \emph{c}-\propenylcat \cite{hua11a}, 
spectroscopic signals from both isomers have been previously observed in the laboratory, 
and also, both are predicted to exist in space. 

Initial reports on \propenylcat  focused on its structural characterization 
by infrared (IR) spectroscopy in neon matrices and solutions of salts \cite{wys01,cra86}. 
Later, the gas-phase IR photodissociation (IRPD) spectra of weakly bound complexes of \propenylcat 
with several tags, including Ne, Ar, H$_2$, O$_2$, N$_2$ and CO$_2$ have been reported both 
experimentally and theoretically \cite{rot02,dop02c,dop02d,bot11,bot11a,bot11b,mar20}. 
Additional knowledge on the isomers of \propenylcat comes from photoelectron \cite{gao12,gar18} and electronic \cite{gar18} spectroscopic studies. 
While the vibrational frequencies obtained from IRPD spectra of such complexes provide useful 
structural information, astronomical identification often requires high resolution studies of the bare species. 
Such information was first provided by Zhao et al. \cite{zha14}, who investigated the 
rovibrational spectrum of the $\nu_4$ antisymmetric C-H stretch of \emph{c}-\propenylcat. 
Their spectral identification was supported by  high level quantum chemical calculations \cite{hua11a,hua11}. 
But due to the absence of a permanent dipole moment in \emph{c}-\propenylcat, 
its detection by radio astronomy is prohibited.

With an eye on a potential radio astronomical detection, 
we investigated those isomers and isotopologues of \propenylcat which possess a permanent dipole moment,
and   studied  the rovibrational and rotational spectra of the singly deuterated 
c-C$_3$H$_2$D$^+$ \cite{gupta23} and the open chain \PI cation.
These investigations culminated in the detection of \PI in the cold molecular cloud TMC-1, 
whereas only upper limits could be given for the column density of  c-C$_3$H$_2$D$^+$ in the same source \cite{silva23}.
In this work, we would like to give a thorough account on the rovibrational spectroscopy of 
\PI, for which the $\nu_1$ and $\nu_3$+$\nu_5$ bands have been measured in high resolution. 
A brief description of the double resonance scheme, which  enabled the measurement of pure rotational 
lines, is also given here.


\section{Experimental}

The rovibrational spectrum of \PI was recorded in the 4~K 22-pole ion 
trap instrument COLTRAP, which has been described in detail previously \cite{asv10,asv14}. 
The \PI ions were created in a storage ion source (SIS) via electron 
impact ionization ($E_e \approx 25-30$~eV) of the precursor gas allene (C$_3$H$_4$), 
which was admitted to the source at a pressure of about $10^{-5}$~mbar. 
The created ions were then extracted from the source, 
mass selected in a first mass filter for mass 39~u, which corresponds to all possible isomers of C$_3$H$_3^+$ produced in the SIS and other possible species with the same mass, 
and injected into the trap. The trap is mounted on a Sumitomo cold head and maintains 
an ambient temperature of 4~K. To ensure thermalization of the ions to the 
cryogenic temperature, the trap is constantly filled with He 
buffer gas ($n$ $\approx$ 10$^{13}$ cm$^{-3}$).

Once trapped, the rovibrational transitions of \PI were measured using the 
novel leak-out spectroscopy (LOS) method \cite{scm22a}. 
LOS is a universal action spectroscopic approach, 
which is based on the escape of ions from the trap due to collision-induced 
transfer of (a fraction of) the ion's internal energy into kinetic energy. 
Specifically, after the ions are loaded into the trap, they are cooled for 40~ms  
 via collisions with He, 
 after which they are irradiated for 500~ms by an IR beam traversing the trap. 
In the case of resonance, the ions are elevated into vibrationally excited states.
These vibrationally excited ions can then collide with noble gas atoms 
present in the trap. We used here Ne as the collision partner and it was pulsed additionally into the trap as a 1:3 Ne:He mixture via a piezoelectrically actuated valve. 
When an ion encounters a Ne atom in the trap, 
a substantial part of  the vibrational energy can be transferred 
into kinetic energy of the colliding pair.
With a  neutral-to-ion mass ratio of 20:39   
 a maximum kinetic energy of $[20/(20+39)] \cdot 3200$~cm$^{-1}$= 1085~cm$^{-1}$ $\approx 0.13$ eV
can be gained by the ion. 
By keeping the electrostatic potential barrier at the exit electrode of the 
trap  below 130~mV, fast \PI ions may leak out from the trap. 
These  escaping ions are then mass selected in a second mass filter, 
and counted in an ion detector. By repeating these trapping cycles 
at a rate of 1~Hz, and  counting the escaping ions as a 
function of the laser wavenumber, 
the rovibrational spectrum of \PI is recorded.

The IR beam was supplied by a continuous wave optical parametric oscillator 
(cw-OPO, Aculight Argos Model 2400, Module C) operating in the 3 $\mu$m spectral region. 
The beam entered the vacuum environment through a diamond window (0.6~mm thickness, Diamond Materials GmbH), irradiated the ions by crossing the 22-pole trap, and exited the instrument via a CaF$_2$ window, 
after which it was absorbed by a power meter. 
The measured power was on the order of a few hundred mW. 
The frequency of the IR radiation was measured by a Bristol wavemeter (Model 621A), which has an accuracy of about 0.001~cm$^{-1}$. 

The pure rotational transitions of \PI are recorded by using a double-resonance spectroscopy scheme
based on LOS \cite{asv23}. 
For this, the IR beam is overlapped with a mm-wave beam outside the apparatus, 
and both steered into the 22-pole trap simultaneously. 
The diamond window mentioned above guarantees transparency for both frequency regimes. 
The mm-wave radiation is produced by a microwave synthesizer (Rohde \& Schwarz SMF 100A) 
that drives an amplifier/multiplier chain source (Virginia Diodes Inc. or Radiometer Physics GmbH). 
The synthesizer is locked 
to a Rb atomic clock with a typical stability of $10^{-11}$. 


\section{Results} \label{sec:results}

\subsection{Identification of carrier of spectroscopic signal} \label{sec:spectralanalysis}

\begin{figure} [h] 
 \includegraphics[width=\textwidth]{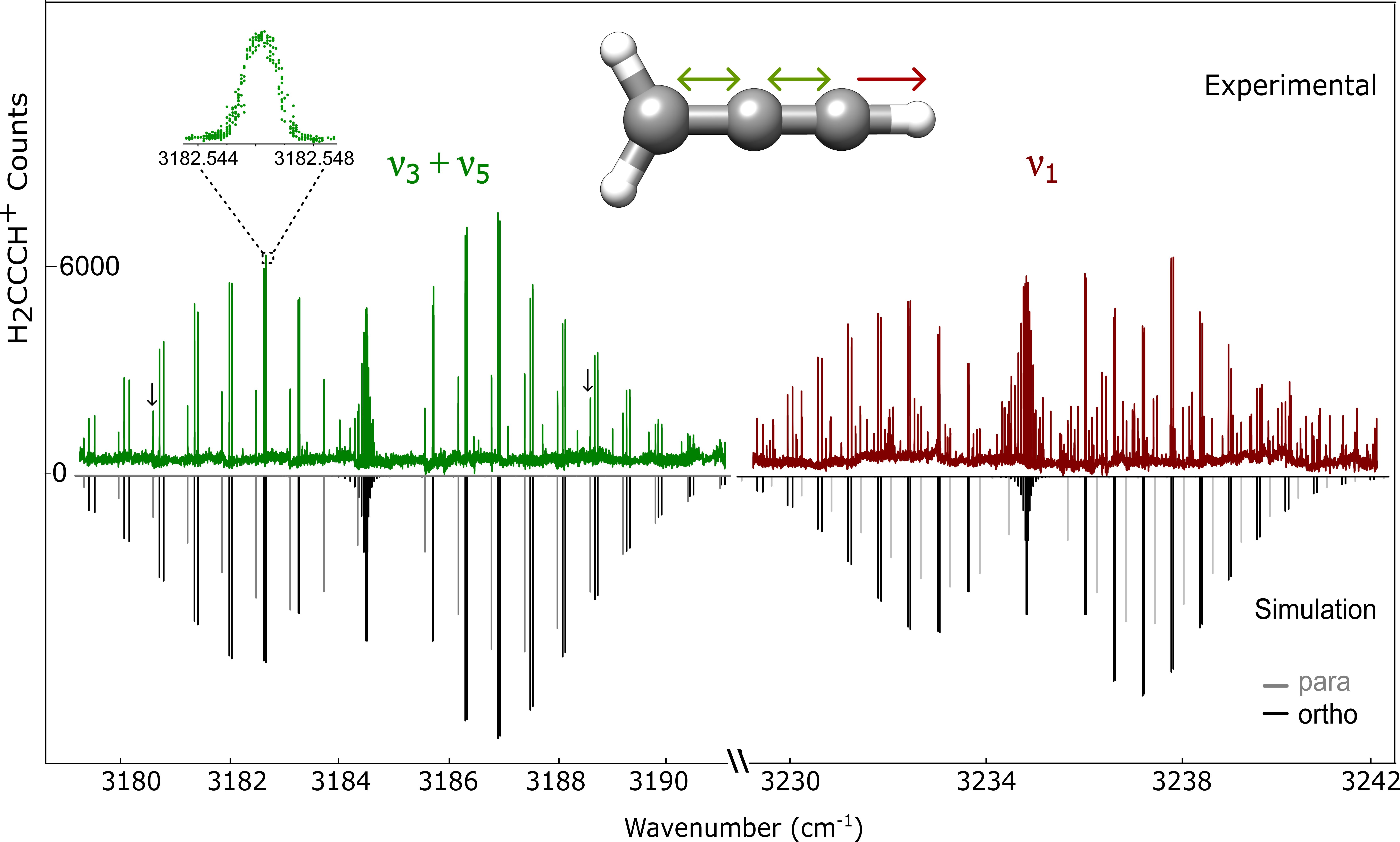}
 \caption{\label{fig:ro-vib} Rovibrational spectra of the $\nu_3+\nu_5$ (3184.915~\wn, green) 
 and $\nu_1$ (3235.334~\wn, red) bands of \PI recorded using the novel LOS method.
 The top portion shows the experimental spectra, 
 while the simulated spectra are given at the bottom.
 The simulations are obtained at a rotational temperature of $T_{rot} = 10$~K,
 using the final fitted parameters given in Table~\ref{tab:fit}. In the simulations, 
 the para ($K_a=0$) and ortho ($K_a=1$) transitions are distinguished by the gray and black colours, respectively, with an ortho-para ratio of 3:1.
 The inset shows the rovibrational line 2$_{12}$ $\leftarrow$ 3$_{13}$ in the $\nu_3+\nu_5$ band. 
 With a FWHM of 42~MHz, the kinetic temperature of the ions is estimated as $T_{kin}=15$~K.}
 \end{figure}

\PI is a closed-shell molecule of point group symmetry $C_{2v}$ with 
an electronic ground state of $^1A_1$. It is a near-prolate asymmetric top with 
an asymmetry parameter $\kappa=-0.9976$, calculated from the rotational constants given by Huang et al.\cite{hua11a}.
The rotationally resolved spectra of the $\nu_1$ and $\nu_3+\nu_5$ parallel bands (both $a_1$ symmetry) are 
depicted in Fig.~\ref{fig:ro-vib} in red and green traces, respectively. 
A simulation of the spectral features using the PGOPHER program \cite{wes17} 
is also provided as black and gray sticks in Fig.~\ref{fig:ro-vib} for illustration. 
The identification of these features as being the $\nu_1$ acetylenic C-H stretch
and the $\nu_3+\nu_5$ combination band of both C-C stretches
is supported by high level quantum chemical calculations \cite{hua11a} and earlier 
observations from IRPD experiments \cite{mar20}.
Also, based on these works, 
we exclude fundamental or combination/overtone bands of 
c-\propenylcat to be responsible for the observed spectral features.
Specifically, measurements of these particular bands 
from Ne-tagging experiments  \cite{rot02} locate the $\nu_3+\nu_5$ band at 3184~\wn and the $\nu_1$ band at 3238~\wn. 
The latter is the assignment  given later by Botschwina et al.~\cite{bot11}. Ar-tagging experiments \cite{ric10} 
observe the $\nu_3+\nu_5$ band at 3182~\wn, also re-assigned by Botschwina et al. \cite{bot11}, and the $\nu_1$ band at 3238~\wn. 
These assignments are also in agreement with the high level \emph{ab initio} 
calculations from Huang et al. \cite{hua11a}, namely $\nu_3+\nu_5$ and $\nu_1$ at 3193 and 3228~\wn, respectively. 
In a more recent Ne-tagging experiment, Marimuthu et al. \cite{mar20} calculate the $\nu_1$ band 
to be at 3230~\wn, but neither $\nu_1$ nor $\nu_3+\nu_5$ were observed in 
their Ne-C$_3$H$_3^+$ IRPD spectrum (see their Fig.~4).

%


\subsection{Spectral Analysis}

With the lines of both bands in Fig.~\ref{fig:ro-vib} being  \emph{a}-type transitions 
(i.e. $\Delta K_a =0, \Delta K_c = \pm 1$), 
we anticipated observing transitions with
$K_a = 1 \leftarrow 1$  and $K_a = 0 \leftarrow 0$,  
whereas transitions with $K_a \ge 2$ were not expected due to the 
large $A$ rotational constant of \PI and the low temperature conditions in the trap.
The most dominant features in the spectra were the series of $K_a = 1 \leftarrow 1$ doublets, 
which were readily identified in both  bands reported here. Preliminary fits of the rotational 
constants $B$ and $C$ from these observations agreed well with those 
predicted by high level \emph{ab initio} calculations  \cite{hua11a}, confirming further the unequivocal detection of \PI.
Although peaks with $K_a = 0 \leftarrow 0$ were expected to be relatively intense,  
it was very challenging to identify these transitions in the $\nu_1$ band because these lines  (i) turned out to be somewhat weaker
(please see Fig.~\ref{fig:ro-vib} and discussion in Section~\ref{sec:discuss}),   
(ii) exhibited  perturbations, (iii) were embedded in   
a plethora of unidentified spectral features (Fig.~\ref{fig:ro-vib}),
and,  in hindsight, (iv) the calculated $A_1$ rotational constant (i.e., the rotation-vibration interaction term $\alpha_1^A$)
turned out to be misleading in sign and magnitude, see Table~\ref{tab:fit}.

Fortunately, the combination band $\nu_3+\nu_5$ contained only few unidentified lines, 
and exhibited only small perturbations, 
allowing the $K_a = 0 \leftarrow 0$ transitions to be readily identified. 
The assignment of the  $K_a = 0 \leftarrow 0$ subband within $\nu_3+\nu_5$
was helped by a small perturbation
in the upper $J_{KaKc}=6_{06}$ quantum level,  
leading to  a splitting of 0.064~\wn which could be spotted in the corresponding
transitions in the P- and R-branches and are indicated by small arrows in Fig.~\ref{fig:ro-vib}. The split lines are also given in the line lists in the supplementary material. 
These assignments allowed first measurements of
pure rotational lines with $K_a = 0 \leftarrow 0$ using double resonance spectroscopy, as
outlined in the following section, and thus a complete determination of the 
ground vibrational state of \PI was obtained.

Due to the difficulties mentioned above, the search for the  $K_a = 0 \leftarrow 0$ subband of $\nu_1$ was  more challenging.  
To identify these transitions, we used our knowledge gained of the ground state rotational structure, and 
compared it to  combination differences (CDs, see, e.g., references \cite{joh92,asv15,bra17}) formed blindly in the $\nu_1$ state.
First, we removed all assigned $K_a = 1 \leftarrow 1$ transitions from 
the $\nu_1$ spectrum and formed all possible CDs with the remaining lines. 
With about $N=130$ unassigned lines, we obtained $N(N-1)/2=8385$ CDs. 
This list of CDs must contain those of the 
ground state, the so-called GSCDs, with $K_a = 0$. 
From our mm-wave measurements, these  GSCDs are constrained to kHz accuracy. 
Based on this knowledge, one can select pairs of lines from all possible CDs, 
one in the P-branch and one in the R-branch, namely $P(J+2)$ and $R(J)$, 
which are connected by a common rotational quantum level $J+1$ in the upper $\nu_1$ state 
and form the GSCD $\approx B_0 (4J+6)$. 
These pairs of lines with the proper GSCD are then potential members of the $K_a = 0 \leftarrow 0$ subband. 
This approach is illustrated by a table of GSCDs in the supplementary material.
Our final assignments of the $K_a = 0 \leftarrow 0$ lines in $\nu_1$ were 
then confirmed also with double resonance spectroscopy measurements, as shown in the following subsection.

A detailed list  of all assigned lines within $\nu_1$ and $\nu_3+\nu_5$ is given in the supplementary material.
Overall, the observed rovibrational lines in both bands exhibit narrow Doppler widths, 
given the low temperature of the ion trap. 
The line 2$_{12}$ $\leftarrow$ 3$_{13}$ in the $\nu_3+\nu_5$ band depicted in the inset 
of Fig.~\ref{fig:ro-vib}, for example, has a full width at half maximum (FWHM) of about 42~MHz which 
corresponds to a kinetic temperature of 15~K. 
The obtained temperature is slightly higher than the nominal trap 
temperature due to known heating effects \cite{asv09}.


\subsection{Double-resonance rotational measurements}

Rotational-vibrational double resonance methods are a superb tool to measure 
rotational fingerprints of molecular ions investigated in cold ion traps \cite{asv21d}. 
The particular double resonance approach which involves the LOS action 
spectroscopic scheme has only recently been developed \cite{asv23}. 
The application of this rotational method
to \PI, and in particular the subsequent detection of this ion in the cold 
molecular cloud TMC-1 have already been documented \cite{silva23}.
In that publication, the 14 laboratory mm-wave transitions as well as ground state rotational 
and centrifugal parameters are listed for \PI. 
Additionally, a summary of all measured rotational transitions as well as a comparison to a simulation 
is  given in Fig.~S1 of the supplementary material.

\begin{figure}
\includegraphics[width=0.8\textwidth]{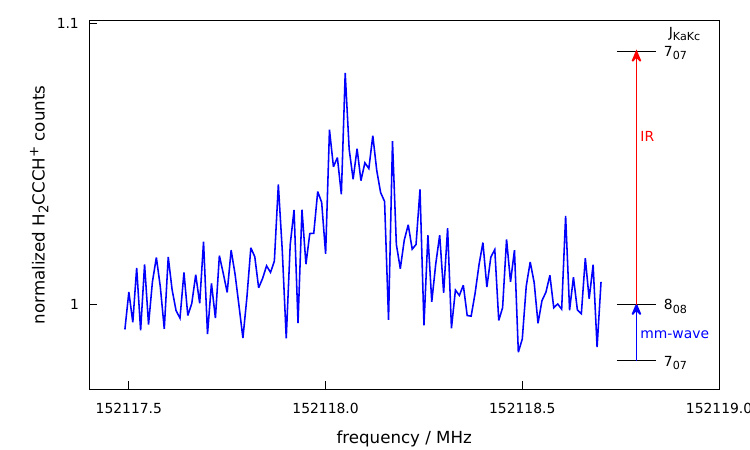}
\caption{\label{fig:rot} 
Pure rotational transition $J'_{Ka'Kc'}$ $\leftarrow$ $J"_{Ka"Kc"}$ = $8_{08}$ $\leftarrow$ $7_{07}$ 
(blue trace and blue arrow)
of \PI,  recorded using the double resonance spectroscopic scheme. 
For this measurement,  the IR laser frequency (red arrow)
was kept fixed on resonance with the $\nu_1$ $7_{07}$ $\leftarrow$ $8_{08}$ rovibrational transition at 3230.208~\wn,
while the mm-wave frequency was stepped in an up-and-down manner in 10~kHz steps.}
\end{figure}

In this paper, we would like to highlight the ability of the double resonance method
to establish the connectivity of transitions and thus to confirm assignments.
%
%
As outlined above, these assignment checks were crucial
for the  transitions with $K_a = 0 \leftarrow 0$, in particular within the heavily  perturbed $\nu_1$ band.
As an example, Fig.~\ref{fig:rot} shows the  rotational line $8_{08}  \leftarrow  7_{07}$ at 152118.074~MHz,
detected with the IR laser fixed at  3230.208~\wn,  the $7_{07}  \leftarrow  8_{08}$ 
transition  in the $\nu_1$ band. 


\subsection{Spectroscopic fits} \label{sec:fit}

The spectroscopic constants for the $\nu_1$ and $\nu_3+\nu_5$ bands were obtained by fitting the assigned lines using a 
Watson's S-reduced Hamiltonian in the \emph{I$^r$} representation as implemented in Western's PGOPHER program \cite{wes17}. 
In these fits, the values of the ground state rotational and centrifugal distortion constants were kept fixed to those 
determined from pure rotational transitions due to their higher accuracy \cite{silva23}. 
As expected, the ground state parameters of \PI  are very similar to those of its deprotonated counterpart, H$_2$CCC~\cite{vrt90},
or those of ketene, H$_2$CCO \cite{joh72,joh92}.
Such molecules are distinguished by their large $A$ rotational and $D_K$ centrifugal parameters.
As the centrifugal distortion constants in the vibrationally excited states could not be well determined, 
they were also fixed to their ground state values, see Table~\ref{tab:fit}.
For the $\nu_1$ band, 54 assigned lines were used in the fit, 49 of which are from the $K_a = 1 \leftarrow 1$ subband.
Although many transitions with $K_a = 0 \leftarrow 0$ could be identified in the $\nu_1$ state via combination differences
(see the complete line list in the supplementary material), 
only five were used in the final fit due to strong perturbations (shifts and splittings). 
For  the $\nu_3+\nu_5$ band, 65 lines were assigned (see supplementary material), 
of which  62 lines within the  $K_a = 0 \leftarrow 0$ 
and $K_a = 1 \leftarrow 1$ bands were used in the final fit. 
It is worth noting that to obtain a satisfactory fit of the 62 lines in the $\nu_3+\nu_5$ band, 
the $D_{JK}$ constant had to be released and its derived value (18.2(1)~MHz) deviates 
significantly to that of the ground state (0.4388~MHz). Thus, the value of $D_{JK}$ presented here 
should be taken with caution as this is likely an artefact of the fit most probably related to 
perturbations in the $K_a = 0 \leftarrow 0$ subband of $\nu_3+\nu_5$, as exemplified above for 
the transitions with upper $J_{KaKc}=6_{06}$ quantum numbers. In an alternative fitting procedure, 
we kept the $D_{JK}$ constant fixed to the ground state value and only included 
transitions within the $K_a = 0 \leftarrow 0$ subband with J $<$ 5. In this fit, 
a reasonable rms error of the obs-calc values of 0.0019 \wn was obtained with the $K_a = 0 \leftarrow 0$ 
lines exhibiting the largest deviations on the order of 0.006 \wn, in agreement with the observed perturbations. 
Nevertheless, the fitted parameters of the first fit are reproduced in
Table \ref{tab:fit} as this allowed the rotational constants and band origin to be derived with smaller uncertainties.
In the future, the observation of transitions with $K_a \ge 2$  (e.g. experiments at higher temperatures) 
will certainly improve the quality of the fit and permit $D_{JK}$ to be better determined within $\nu_3+\nu_5$.


Overall, the fits for both states presented in Table \ref{tab:fit} are quite reasonable with rms errors of the obs-calc values of 
0.0004~\wn and 0.0003~\wn for $\nu_1$ and $\nu_3+\nu_5$, respectively.  
Please note that the fit results may be somewhat optimistic, as many parameters were fixed.
For instance, the uncertainties in the values of $A_{1}$ and $A_{3+5}$ reflect only that of the fit and do not take into account the large uncertainty of $A_0$ itself, which is 247~MHz \cite{silva23}. 
Future investigations of a b-type perpendicular band of \PI, e.g.\ the antisymmetric CH$_2$ 
stretch at around 3093~\wn \cite{ric10},
may help to constrain $A_0$ and $D_K$ better. 
A similar strategy has been followed for ketene, H$_2$CCO \cite{joh72}, 50 years ago.

\begin{table} 
\begin{center}
\caption{\label{tab:fit} Spectroscopic parameters of \PI (Watson's S-reduced Hamiltonian) 
obtained by fitting  the  rovibrational transitions given in the supplementary material, 
using Western's PGOPHER program \cite{wes17}. 
The uncertainty in the last digits of each parameter is reported in parentheses. 
The units are MHz, unless otherwise noted.}
\begin{tabular}{l r@{}l  r@{}l | r@{}l r@{}l | r@{}l r@{}l  }
\hline
Parameter & \multicolumn{4}{c}{ground}  & \multicolumn{4}{c}{ $\nu_1$}  & \multicolumn{4}{c}{ $\nu_3+\nu_5$} \\
          & \multicolumn{2}{c}{exp $^a$} &  \multicolumn{2}{c}{\it{ab initio} $^b$} &   \multicolumn{2}{c}{this work}  
          & \multicolumn{2}{c}{\it{ab initio} $^b$} &    \multicolumn{2}{c}{this work}       & \multicolumn{2}{c}{\it{ab initio} $^b$}  \\
\hline
 $\nu$ / cm$^{-1}$   & &                   &      0&     & 3235&.3338(2)    & 3239&.0   & 3184&.9154(1)      & 3201&.8\\
  $A$                 &    281856\,&        & 281911&.9   & 274403&(6)      & 281933&.2 & 267652&(4)        &  281534&.8\\
  $B$  	            &   9675.&841        &   9580&.2   & 9637&.9(1)       & 9554&.3   & 9625&.3(1)         &  9501&.2\\
  $C$  	            &   9342.&877        &   9251&.9   & 9325&.6(1)       & 9227&.4   & 9315&.1(1)         &  9175&.9\\
  $D_J\times10{^3}$   &    2.&942          &      3&     & 2&.942 $^c$       & -&        &  2&.942  $^c$          & -& \\
  $D_{JK}$            &    0.&4388         &      0&.479 & 0&.4388 $^c$      & -&        &  18&.2(1)               & -& \\ 
  $D_K$               &    20.&678         &     20&.862 & 20&.678 $^c$      & -&        &  20&.678 $^c$       & -& \\
  $d_1\times10{^3}$   &    -0.&121         &      0&     & -0&.121 $^c$      & -&        &  -0&.121  $^c$         & -& \\
  $d_2\times10{^3}$   &    -0.&051         &      0&     & -0&.051 $^c$      & -&        &  -0&.051 $^c$       & -& \\
\hline
\end{tabular}
\end{center}
$^a$: ground state parameters fixed to those of Silva et al.~\cite{silva23}. $D_K$ and $d_2$ are calculated.  \\
$^b$: values from Huang et al. \cite{hua11a}  \\
$^c$: Fixed to ground state value \\
\end{table}


\section{Discussion and Outlook} \label{sec:discuss}

The successful measurement of the rovibrational spectrum of the fundamental $\nu_1$ band (C-H stretch)  
and the combination band $\nu_3+\nu_5$ (C-C stretches) for the  \PI cation presented here 
illustrates the broad applicability and remarkable sensitivity of the recently developed LOS method. 
The rotational-vibrational double resonance method using LOS, recently demonstrated in our group for the 
HC$_3$O$^+$ ion \cite{asv23}, enabled not only pure rotational transitions for \PI to be obtained but 
also the perturbed $\nu_1$ band to be deciphered. A search in the cold interstellar cloud TMC-1 finally 
led to the detection of this ion in space \cite{silva23}.
A similar approach was followed for another member of the C$_3$H$_3^+$ family,
the singly deuterated c-C$_3$H$_2$D$^+$ cation \cite{gupta23}, 
but unfortunately its low dipole moment and low abundance
inhibited detection in TMC-1, and only an upper limit for its column density could be given \cite{silva23}.
In the future, we expect that LOS will enable the characterization of more cations in the laboratory,
potentially leading to  their  detection in space by radio astronomy.

While these results  are relevant to astronomers, 
interesting insights into the molecular structure of \PI 
upon vibrational excitation are obtained. 
In particular, the value of the rotational constant $A$ (Table~\ref{tab:fit}) decreases 
upon excitation into the $\nu_1$ and $\nu_3+\nu_5$ states, by 2.7~\% and 5~\%, respectively.
This suggests significant structural changes when the terminal C-H or C-C bonds are excited. 
The experimentally obtained vibration-rotation interaction constants are about 
$\alpha^A_1 = 7486$~MHz and $\alpha^A_{3+5} = 14483$~MHz, respectively.
As the backbone of \PI lies along the $a$-axis with only the methylene hydrogens 
pointing away from this axis, the value of $A$ is intimately related to the value of the average $\angle$HCH angle. 
Considering that $A$ is inversely proportional to the moment of inertia component $I_a$, 
it becomes evident that the smaller values of $A_{1}$ and $A_{3+5}$ are 
likely due to the C-H and C-C stretches that move the methylene hydrogens 
further apart by increasing the $\angle$HCH angle when vibrating. 
This effect is more pronounced in the $\nu_{3+5}$ band as the C-C bonds are 
spatially closer to the CH$_2$ group. Although the predictions from Huang et al. \cite{hua11a} 
support the experimental assignments and show overall good agreement with the determined values, 
the theoretical calculations do not capture well the trend observed 
for the changes in the values of $A$ upon vibrational excitation, 
as seen in Table~\ref{tab:fit}. The calculated predictions are $\alpha^A_1 = -21.277$~MHz
and $\alpha^A_{3+5} = 377.099$~MHz.
These discrepancies may be associated with perturbations in \PI, 
which are not well described by standard computational methods. 
Thus, our high-resolution study of \PI also provides data which can 
contribute to the improvement and benchmark of theoretical models.

Although \PI has been well characterized in this work, many open questions remain 
concerning its rovibrational spectroscopy. The above mentioned perturbations 
affect both rovibrational bands, but particularly $\nu_1$ exhibits severe line shifts and
splittings. Such severe perturbations have also been observed for ketene~\cite{joh72,dun87,joh92},
and traced back to the presence of low-lying vibrational states, whose overtones and combination bands
give rise to Fermi and Coriolis interactions~\cite{dun87,joh92}. The unusual effective value for $D_{JK}$ 
determined for $\nu_3+\nu_5$ is certainly a consequence of this,
as well as the intensity pattern observed both for $\nu_1$ and $\nu_3+\nu_5$.
As \PI contains a plane of symmetry with two equivalent hydrogen atoms, \PI has  two nuclear spin states, 
ortho (levels with odd values of $K_a$) and para (levels with even values of $K_a$) 
with an expected ortho-para ratio of 3:1. 
This situation is very similar to, e.g., neutral H$_2$CCC \cite{Cernicharo1991,vrt90}
or the aforementioned  ketene, H$_2$CCO \cite{joh72,joh92}.
From the radio astronomical observations of \PI, in which the para 2$_{02}$-1$_{01}$ 
and ortho 2$_{12}$-1$_{11}$ and 2$_{11}$-1$_{10}$ transitions were seen \cite{silva23}, 
a consistent ortho-para ratio of 3.3($\pm$0.6):1 was derived. 
In the rovibrational spectra in Fig.~\ref{fig:ro-vib}, however, 
the $K_a = 0 \leftarrow 0$ para-transitions exhibit somewhat lower than expected intensities. 
To reproduce the observed IR spectra, 
an 'artificial' ortho-para ratio of roughly 4:1 would be needed. 
We can exclude that the unusual ratio is related to the formation processes of \PI 
by electron impact ionization of allene (C$_3$H$_4$) in our ion source,
because  essentially the same spectrum was obtained using a HCCD/CH$_4$ precursor mixture, 
see Fig.~S3 in the supplementary material.
Also, we do not suspect that our novel LOS method to be responsible for this peculiar observation, as 
experiments with other ions containing two identical protons exhibited the expected nuclear spin statistics,
e.g. C$_2$H$_2^+$ \cite{scl23} and  c-C$_3$H$_2$D$^+$  \cite{gupta23}. 
The intensities observed in our spectrum, however, depend on multiple factors like the transition dipole moment, 
population distribution, trap time, and more. 
So, to determine an accurate ortho-para population ratio of \PI in the ion trap all ortho (or para) species shall be leaked out from the trap and the remaining para (or ortho) fraction shall be measured. 
This new method of determining the nuclear-spin composition in a trap\footnote{Patent pending: DE 10 2021 127 556.3 (Universit\"at zu K\"oln), 22.10.2021} is currently tested in our laboratory for a number of simple cases. Results from these measurements will be published separately.
%
In addition, we plan to perform rovibrational spectroscopy of the b-type antisymmetric CH$_2$ 
stretch or one of the fundamental a-type C-C stretches \cite{bas23} of \PI. 
Measurements at higher trap temperature to populate the $K_a=2$ states
or the rovibrational measurement of the perdeuterated version, D$_2$CCCD$^+$,
are also in reach.\\

\section{Acknowledgements} \label{sec:ack}
The authors want to dedicate this contribution to Prof.~Attila Cs\'asz\'ar on the occasion of his 60$^{th}$ birthday.
The authors also want to thank Xinchuan Huang for reading the manuscript prior to submission.
This work has been supported by an ERC advanced grant (MissIons: 101020583) and via Collaborative Research Centre 1601 
(project ID: 500700252, sub-project C4) funded by the Deutsche Forschungsgemeinschaft (DFG) and DFG SCHL 341/15-1 
(``Cologne Center for Terahertz Spectroscopy''). W.G.D.P.S. thanks the Alexander von Humboldt Foundation for 
support through a postdoctoral fellowship. 
J.L.D. acknowledges the support from the MCINN project PID2020-113084GB-I00/AEI/10.13039/501100011033 and the CSIC project ILINK+ LINKA20353.


\end{document}


\section{Calibration measurements}

We performed additional calibration measurements with neutral C$_2$H$_2$ contained in a gas cell. We measured several lines of C$_2$H$_2$ between 3190 and 3240~cm$^{-1}$ whose frequencies are given in the HITRAN database (J. Quant. Spectrosc. Radiat. Transfer, 2022, 266, 107949). Following this procedure, we shifted our data up by 0.007~cm$^{-1}$. With the calibration, the accuracy of our IR measurements should be on the order of 0.001~cm$^{-1}$.

\section{Linelist of measured transitions and residuals}

\begin{spacing}{.7}
\begin{longtable} [c] {c| c| c} 
\caption{Frequencies and residuals (in \wn) of the rovibrational transitions observed for the $\nu_1$ band of \PI. Lines marked with an asterisk (*) were not included in the fit.}\\
\hline
transition & frequency & obs-calc \\
\hline
$8_{17} \leftarrow 9_{18}$   &  3229.2485 &  -0.0002\\
$8_{18} \leftarrow 9_{19}$   &  3229.3732 &  -0.0002\\
$8_{08} \leftarrow 9_{09}$   &	3229.5255* & \\
$7_{16} \leftarrow 8_{17}$   &  3229.9057 &  -0.0002\\
$7_{17} \leftarrow 8_{18}$   &  3230.0136 &  -0.0004\\
$7_{07} \leftarrow 8_{08}$   &  3230.2076 &  -0.0006\\
$6_{15} \leftarrow 7_{16}$   &  3230.5609 &  -0.0001\\
$6_{16} \leftarrow 7_{17}$   &  3230.6527 &  -0.0005\\
$6_{06} \leftarrow 7_{07}$   &  3230.8424* & \\ 
$5_{14} \leftarrow 6_{15}$   &  3231.2139 &  -0.0000\\
$5_{15} \leftarrow 6_{16}$   &  3231.2904 &  -0.0005\\
$5_{05} \leftarrow 6_{06}$   &  3231.4815* & \\
$4_{13} \leftarrow 5_{14}$   &  3231.8646 &  -0.0002\\
$4_{14} \leftarrow 5_{15}$   &  3231.9266 &  -0.0005\\
$4_{04} \leftarrow 5_{05}$   &  3232.1126* & \\
							 &	3232.1915* & \\
$3_{12} \leftarrow 4_{13}$   &  3232.5131 &  -0.0003\\
$3_{13} \leftarrow 4_{14}$   &  3232.5615 &  -0.0004\\
$3_{03} \leftarrow 4_{04}$   &  3232.7216* & \\
							 &  3232.7986* & \\ 
$2_{11} \leftarrow 3_{12}$   &  3233.1592 &  -0.0006\\
$2_{12} \leftarrow 3_{13}$   &  3233.1945 &  -0.0007\\
$2_{02} \leftarrow 3_{03}$   &  3233.4284* & \\
$1_{10} \leftarrow 2_{11}$   &  3233.8036 &  -0.0005\\
$1_{11} \leftarrow 2_{12}$   &  3233.8264 &  -0.0006\\
$1_{01} \leftarrow 2_{02}$   &  3234.0636 &   0.0004\\
$8_{18} \leftarrow 8_{17}$   &  3234.6321 &  -0.0003\\
$0_{00} \leftarrow 1_{01}$   &  3234.6988 &  -0.0006\\
$7_{17} \leftarrow 7_{16}$   &  3234.7334 &   0.0002\\
$6_{16} \leftarrow 6_{15}$   &  3234.8214 &  -0.0001\\
$5_{15} \leftarrow 5_{14}$   &  3234.8971 &  -0.0000\\
$4_{14} \leftarrow 4_{13}$   &  3234.9602 &   0.0001\\
$3_{13} \leftarrow 3_{12}$   &  3235.0103 &  -0.0002\\
$2_{12} \leftarrow 2_{11}$   &  3235.0482 &  -0.0001\\
$1_{11} \leftarrow 1_{10}$   &  3235.0734 &  -0.0002\\
$1_{10} \leftarrow 1_{11}$   &  3235.0951 &   0.0000\\
$2_{11} \leftarrow 2_{12}$   &  3235.1129 &  -0.0000\\
$3_{12} \leftarrow 3_{13}$   &  3235.1398 &   0.0001\\
$4_{13} \leftarrow 4_{14}$   &  3235.1756 &   0.0003\\                             
$5_{14} \leftarrow 5_{15}$   &  3235.2203 &   0.0004\\                             
$6_{15} \leftarrow 6_{16}$   &  3235.2738 &   0.0004\\                              
$7_{16} \leftarrow 7_{17}$   &  3235.3362 &   0.0004\\
$8_{17} \leftarrow 8_{18}$   &  3235.4076 &   0.0004\\
$1_{01} \leftarrow 0_{00}$   &  3235.9674 &   0.0010\\
$2_{12} \leftarrow 1_{11}$   &  3236.3393 &   0.0001\\
$2_{11} \leftarrow 1_{10}$   &  3236.3596 &   0.0001\\
$2_{02} \leftarrow 1_{01}$   &  3236.6015* & \\
$3_{13} \leftarrow 2_{12}$   &  3236.9639 &   0.0003\\
$3_{12} \leftarrow 2_{11}$   &  3236.9932 &   0.0004\\
$3_{03} \leftarrow 2_{02}$   &  3237.1630* & \\
							 &	3237.2398* & \\
$4_{14} \leftarrow 3_{13}$   &  3237.5868 &   0.0005\\
$4_{13} \leftarrow 3_{12}$   &  3237.6245 &   0.0006\\
$4_{04} \leftarrow 3_{03}$   &  3237.8225* & \\
							 &	3237.9018* & \\
$5_{15} \leftarrow 4_{14}$   &  3238.2081 &  0.0005\\
$5_{14} \leftarrow 4_{13}$   &  3238.2536 &  0.0008 \\
$5_{05} \leftarrow 4_{04}$   &	3238.4599* & \\
$6_{16} \leftarrow 5_{15}$   &  3238.8275 &  0.0002\\
$6_{15} \leftarrow 5_{14}$   &  3238.8800 &  0.0005\\
$6_{06} \leftarrow 5_{05}$   &  3239.0889* & \\
$7_{17} \leftarrow 6_{16}$   &  3239.4456 &   0.0000\\
$7_{16} \leftarrow 6_{15}$   &  3239.5045 &   0.0004\\
$7_{07} \leftarrow 6_{06}$   &  3239.7222 &  -0.0001\\
$8_{18} \leftarrow 7_{17}$   &  3240.0628 &   0.0005\\
$8_{17} \leftarrow 7_{16}$   &  3240.1266 &   0.0003\\
$8_{08} \leftarrow 7_{07}$   &  3240.3082* & \\
$9_{19} \leftarrow 8_{18}$   &  3240.6781 &   0.0007\\
$9_{18} \leftarrow 8_{17}$   &  3240.7461 &  -0.0003\\
$10_{110} \leftarrow 9_{19}$ &  3241.2885* & \\
$10_{19} \leftarrow 9_{18}$  &  3241.3633 &  -0.0009\\
$11_{111} \leftarrow 10_{110}$ & 3241.9055* & \\
$11_{110} \leftarrow 10_{19}$ &  3241.9782* & \\
\hline
\end{longtable}
\end{spacing}


\begin{spacing}{.7}
\begin{longtable}{l | c | c }
\caption{\label{linelist_nu3+5} Rovibrational transition frequencies and residuals of lines observed in $\nu_{3+5}$. Lines marked with an asterisk (*) were not included in the fit.}\\
\hline
transition & frequency & obs-calc \\
\hline
$8_{08} \leftarrow 9_{09}$   &  3179.1134 &  -0.0000\\
$7_{16} \leftarrow 8_{17}$   &  3179.2071 &   0.0002\\
$7_{17} \leftarrow 8_{18}$   &  3179.3170 &  -0.0001\\
$7_{07} \leftarrow 8_{08}$   &  3179.7685 &   0.0003\\
$6_{15} \leftarrow 7_{16}$   &  3179.8762 &   0.0002\\
$6_{16} \leftarrow 7_{17}$   &  3179.9698 &   0.0001\\
$6_{06} \leftarrow 7_{07}$   &  3180.4156* &  \\
                             &  3180.4220* &  \\
$5_{14} \leftarrow 6_{15}$   &  3180.5410 &   0.0001 \\
$5_{15} \leftarrow 6_{16}$   &  3180.6189 &   0.0000\\
$5_{05} \leftarrow 6_{06}$   &  3181.0709 &   0.0006 \\
$4_{13} \leftarrow 5_{14}$   &  3181.2018 &   0.0002\\
$4_{14} \leftarrow 5_{15}$   &  3181.2646 &  -0.0002 \\
$4_{04} \leftarrow 5_{05}$   &  3181.7181 &   0.0005\\
$3_{12} \leftarrow 4_{13}$   &  3181.8582 &   0.0000 \\
$3_{13} \leftarrow 4_{14}$   &  3181.9071 &  -0.0001\\
$3_{03} \leftarrow 4_{04}$   &  3182.3623 &  -0.0001 \\
$2_{11} \leftarrow 3_{12}$   &  3182.5108 &   0.0001\\
$2_{12} \leftarrow 3_{13}$   &  3182.5462 &  -0.0001 \\
$2_{02} \leftarrow 3_{03}$   &  3183.0044 &  -0.0001\\
$1_{10} \leftarrow 2_{11}$   &  3183.1590 &   0.0001 \\
$1_{11} \leftarrow 2_{12}$   &  3183.1820 &   0.0001\\
$1_{01} \leftarrow 2_{02}$   &  3183.6435 &  -0.0005 \\
$7_{17} \leftarrow 7_{16}$   &  3184.0358 &  -0.0005\\
$6_{16} \leftarrow 6_{15}$   &  3184.1375 &  -0.0003 \\
$5_{15} \leftarrow 5_{14}$   &  3184.2246 &  -0.0004\\
$0_{00} \leftarrow 1_{01}$   &  3184.2802 &  -0.0009 \\
$4_{14} \leftarrow 4_{13}$   &  3184.2976 &  -0.0001\\
$3_{13} \leftarrow 3_{12}$   &  3184.3558 &  -0.0000 \\
$2_{12} \leftarrow 2_{11}$   &  3184.3993 &  -0.0000\\
$1_{11} \leftarrow 1_{10}$   &  3184.4285 &   0.0001 \\
$1_{10} \leftarrow 1_{11}$   &  3184.4499 &   0.0000\\                             
$2_{11} \leftarrow 2_{12}$   &  3184.4638 &   0.0001 \\                             
$3_{12} \leftarrow 3_{13}$   &  3184.4845 &  -0.0000\\                              
$4_{13} \leftarrow 4_{14}$   &  3184.5122 &   0.0000 \\
$5_{14} \leftarrow 5_{15}$   &  3184.5467 &  -0.0001\\
$6_{15} \leftarrow 6_{16}$   &  3184.5881 &  -0.0003 \\
$7_{16} \leftarrow 7_{17}$   &  3184.6366 &  -0.0003\\
$1_{01} \leftarrow 0_{00}$   &  3185.5466 &  -0.0006 \\
$2_{12} \leftarrow 1_{11}$   &  3185.6904 &   0.0001\\
$2_{11} \leftarrow 1_{10}$   &  3185.7103 &   0.0001 \\
$2_{02} \leftarrow 1_{01}$   &  3186.1764 &  -0.0000\\
$3_{13} \leftarrow 2_{12}$   &  3186.3090 &   0.0002 \\
$3_{12} \leftarrow 2_{11}$   &  3186.3377 &   0.0001\\
$3_{03} \leftarrow 2_{02}$   &  3186.8030 &   0.0001 \\
$4_{14} \leftarrow 3_{13}$   &  3186.9241 &   0.0002\\
$4_{13} \leftarrow 3_{12}$   &  3186.9609 &   0.0001 \\
$4_{04} \leftarrow 3_{03}$   &  3187.4275 &   0.0007\\
$5_{15} \leftarrow 4_{14}$   &  3187.5353 &  -0.0003 \\
$5_{14} \leftarrow 4_{13}$   &  3187.5797 &  -0.0001\\
$5_{05} \leftarrow 4_{04}$   &  3188.0487 &   0.0007 \\
$6_{16} \leftarrow 5_{15}$   &  3188.1437 &  -0.0001\\
$6_{15} \leftarrow 5_{14}$   &  3188.1943 &  -0.0002 \\
$6_{06} \leftarrow 5_{05}$   &  3188.6618* &  \\
                             &  3188.6681* &  \\
$7_{17} \leftarrow 6_{16}$   &  3188.7487 &   0.0001 \\
$7_{16} \leftarrow 6_{15}$   &  3188.8051 &   0.0001\\
$7_{07} \leftarrow 6_{06}$   &  3189.2828 &   0.0003 \\
$8_{18} \leftarrow 7_{17}$   &  3189.3499 &   0.0000\\
$8_{17} \leftarrow 7_{16}$   &  3189.4115 &   0.0001 \\
$8_{08} \leftarrow 7_{07}$   &  3189.8954 &  -0.0003\\
$9_{19} \leftarrow 8_{18}$   &  3189.9484 &   0.0006 \\
$9_{18} \leftarrow 8_{17}$   &  3190.0136 &   0.0000\\
$9_{09} \leftarrow 8_{08}$   &  3190.5054 &  -0.0008\\
$10_{110} \leftarrow 9_{19}$ &  3190.5481* &  \\
$10_{19} \leftarrow 9_{18}$  &  3190.6114 &   -0.0001\\
$11_{110} \leftarrow 10_{19}$ &  3191.2052 &   0.0001\\
\hline
\end{longtable}
\end{spacing}

\newpage

\section{Combination differences in the $\nu_1$ $K_a = 0 \leftarrow 0$ subband}

As mentioned in the main manuscript, only after measuring and assigning the $K_a = 0 \leftarrow 0$
lines in the $\nu_3+ \nu_5$ band, it was possible to predict and subsequently measure 
a-type rotational transitions with $K_a = 0 \leftarrow 0$, 
and thus to fully determine the ground state rotational structure of \PI. 
This, in turn, allowed the $K_a = 0 \leftarrow 0$ subband in $\nu_1$ to be identified via ground state combination differences (GSCDs), as shown by the table below.

\begin{table} [H]
\begin{center}
\caption{\label{rotlines} 
GSCDs within $\nu_1$ $K_a = 0 \leftarrow 0$ and comparison to 
data determined by mm-wave measurements. All values in \wn. 
The GSCDs determined via the IR data ("trans1-trans2") are accurate within 0.001~\wn, while the values based on rotational spectroscopy ("mm-wave") are accurate within a few kHz.}
\begin{tabular}{l | l | l | r | c }
\hline
IR tansition2               & IR transition1 & trans2 - trans1 & mm-wave & comment\\
\hline
3235.9674 $1_{01} \leftarrow 0_{00}$ & 3234.0636 $1_{01} \leftarrow 2_{02}$  & 1.9038 $2_{02} - 0_{00}$  &  1.9031750 &\\
3236.6015 $2_{02} \leftarrow 1_{01}$ & 3233.4284 $2_{02} \leftarrow 3_{03}$  & 3.1730 $3_{03} - 1_{01}$  &  3.1719165 & \\
3237.2398 $3_{03} \leftarrow 2_{02}$ & 3232.7986 $3_{03} \leftarrow 4_{04}$  & 4.4412 $4_{04} - 2_{02}$  &  4.4405952 & perturbed/split\\
3237.1630                            & 3232.7216                             & 4.4414                    &            & perturbed/split \\
3237.8225 $4_{04} \leftarrow 3_{03}$ & 3232.1126 $4_{04} \leftarrow 5_{05}$  & 5.7099 $5_{05} - 3_{03}$  &  5.7091859 & perturbed/split\\
3237.9018                            & 3232.1915                             & 5.7103                    &            &  perturbed/split\\
3238.4599 $5_{05} \leftarrow 4_{04}$ & 3231.4815 $5_{05} \leftarrow 6_{06}$  & 6.9785 $6_{06} - 4_{04}$  &  6.9776637 & \\
3239.0889 $6_{06} \leftarrow 5_{05}$ & 3230.8424 $6_{06} \leftarrow 7_{07}$  & 8.2465 $7_{07} - 5_{05}$  &  8.2460036 & \\
3239.7222 $7_{07} \leftarrow 6_{06}$ & 3230.2076 $7_{07} \leftarrow 8_{08}$  & 9.5147 $8_{08} - 6_{06}$  &  9.5141806 & \\
3240.3082 $8_{08} \leftarrow 7_{07}$ & 3229.5255 $8_{08} \leftarrow 9_{09}$  & 10.7827 $9_{09} - 7_{07}$  &  10.7821699 & \\
\hline                                                  
\end{tabular}                                      
\end{center}                                       
\end{table}


\newpage

\section{Ground state rotational transitions and energy levels}

\begin{figure} [H]
    \includegraphics[width=0.6\textwidth]{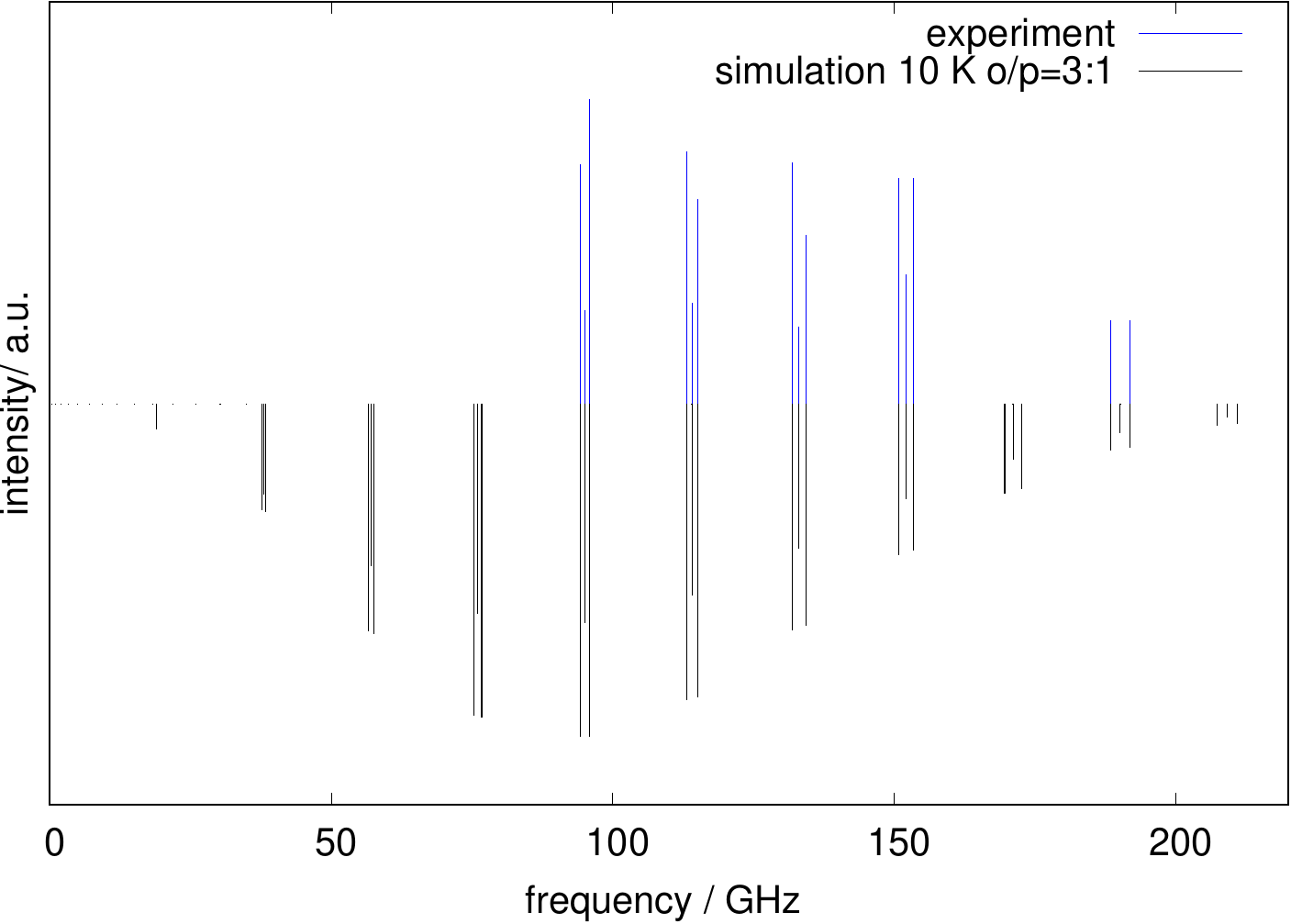}
    \caption{Summary of the 14 rotational lines measured in the course of this project (blue sticks).
    For the simulation using PGOPHER (black sticks), the ortho and para transitions have been simulated 
    separately at a temperature of 10~K, and then plotted with corresponding weights of 3 and 1, respectively.
    }
\end{figure}

\begin{figure} [H]
    \includegraphics[width=0.6\textwidth]{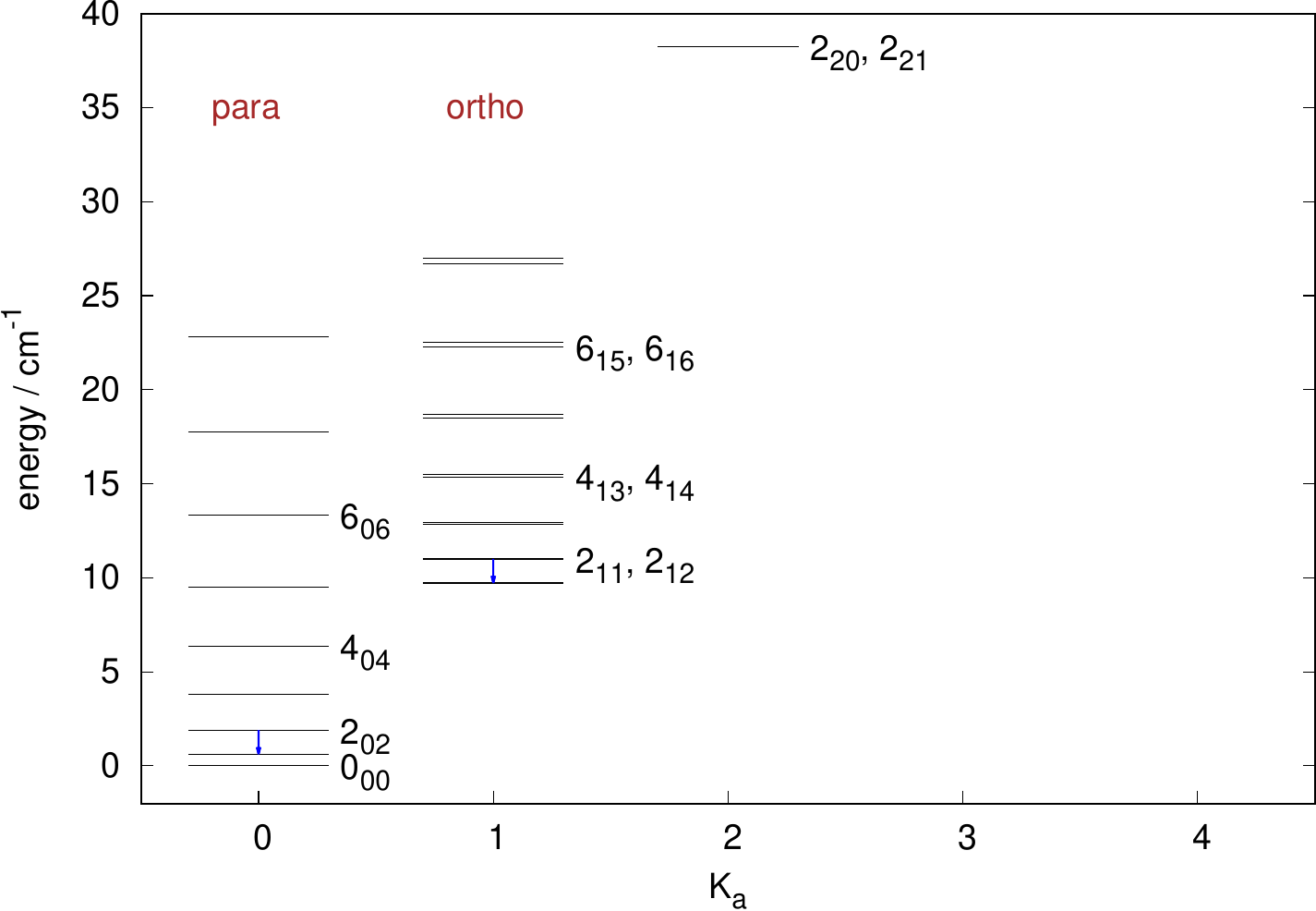}
    \caption{Ground state energy levels of \PI. In our low-temperature experiment,
    only para states with $K_a=0$ and ortho states with $K_a=1$ are populated. 
    }
\end{figure}


\newpage

\section{Comparison of IR spectra from different precursors}

Prior to investigating \PI, we studied in Cologne the IR spectrum of c-C$_3$H$_2$D$^+$, 
produced in the source from the electron impact ionization of a HCCD/CH$_4$ 
mixture (Faraday Discuss. 245, 298–308 (2023), doi: 10.1039/d3fd00068k). 
As the ionization of this mixture also produces substantial amounts of C$_3$H$_3^+$ isomers with mass 39~u, 
our first search for the $\nu_1$ band of \PI\ was carried out with 
this  precursor mixture (shown below in red). 
The observed features were consistent with those predicted for a near prolate asymmetric top, like \PI. 
Since our mass selection could also introduce deuterated species with the same mass (39 u) into the trap, 
we later re-measured the same spectral region using a precursor which does not contain deuterium,  
and we chose allene (C$_3$H$_4$).
This measurement corroborated  \PI\ as the carrier of the spectroscopic signals (shown below in black).

\begin{figure} [H]
    \includegraphics[width=0.8\textwidth]{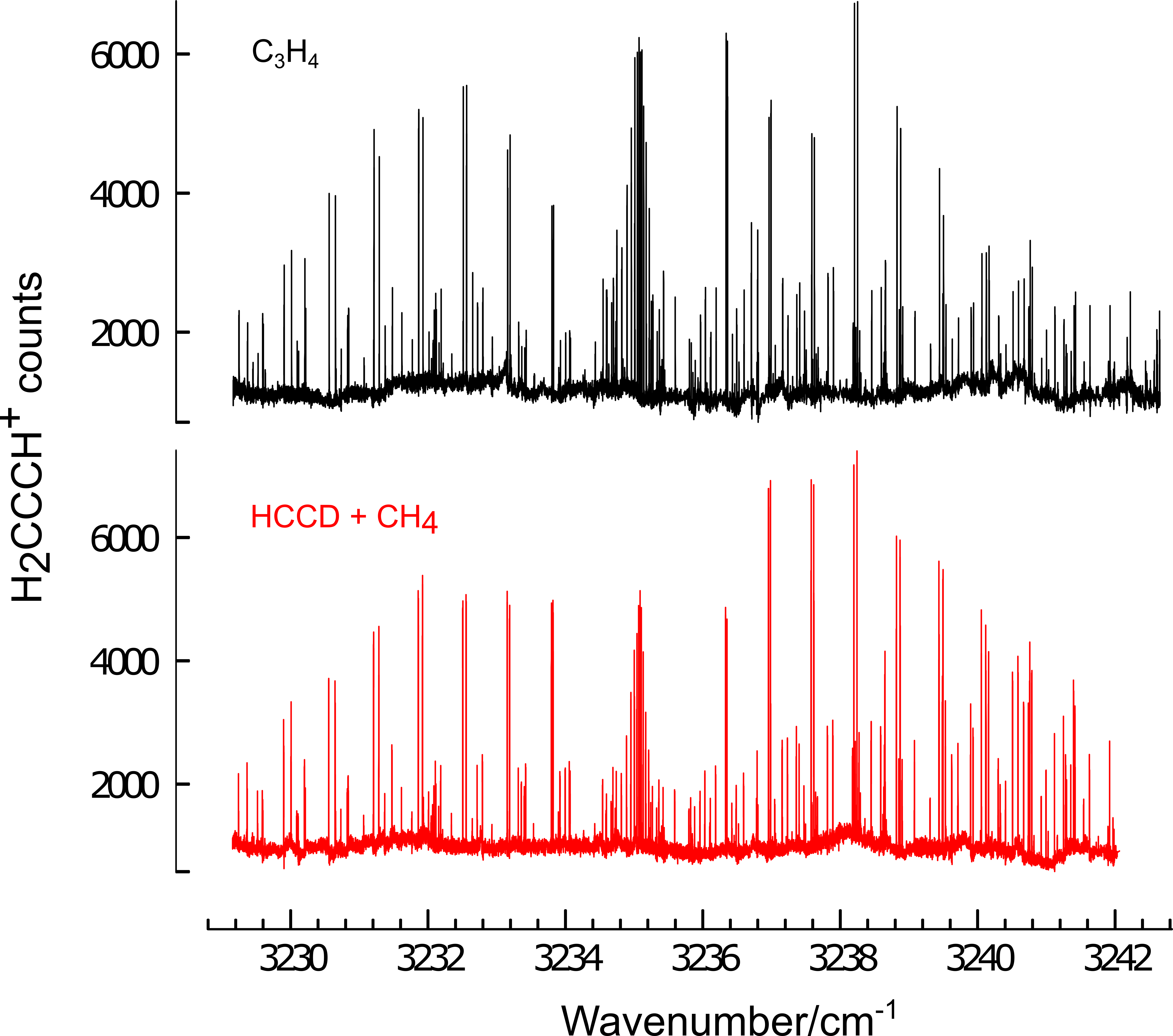}
    \caption{Comparison between the $\nu_1$ band of \PI\  recorded with  
    allene C$_3$H$_4$ (top, black) and with a mixture of HCCD/CH$_4$ (bottom, red) as precursors.}
\end{figure}